\def\<{\langle}
\def\>{\rangle}

\newcommand{\ra}{\;\raise1.0pt\hbox{$'$}\hskip-6pt\partial\;}
\newcommand{\lo}{\;\overline{\raise1.0pt\hbox{$'$}\hskip-6pt\partial}\;}

\newcommand{\degree}{^\circ}

\newcommand{\Abs}{\abstract}
\newcommand{\Ack}[1]{\begin{acknowledgements} #1 \end{acknowledgements}}
\newcommand{\mktt}{\maketitle}

\documentclass{aa}

\usepackage{graphicx,epsfig,natbib,color,times,bm,amsmath,multirow,colortbl,xcolor}

\begin{document}

\title{Fingerprint of Galactic Loop I on polarized microwave foregrounds}

\author{
Hao Liu \inst{\ref{NBI}\,\ref{IHEP}}
}
\institute{
The Niels Bohr Institute \& Discovery Center, Blegdamsvej 17, DK-2100 Copenhagen, Denmark \label{NBI} \and
Key laboratory of Particle and Astrophysics, Institute of High Energy Physics, CAS, 19B YuQuan Road, Beijing, China \label{IHEP}
\\ \email{
\\ liuhao@nbi.dk
}}

\Abs{

\textbf{Context}: Currently, detection of the primordial gravitational waves
using the B-mode of cosmic microwave background (CMB) is primarily limited by
our knowledge of the polarized microwave foreground emissions. Improvements of
the foreground analysis are therefore necessary. As revealed
in~\cite{2018arXiv180410382L}, the E-mode and B-mode of the polarized
foreground have noticeably different properties, both in morphology and
frequency spectrum, suggesting that they arise from different physical
processes, and need to be studied separately.\\

\textbf{Aims}: I study the polarized emission from Galactic loops, especially
Loop I, and mainly focus on the following questions: Does the polarized loop
emission contribute predominantly to the E-mode or B-mode? In which frequency
bands and in which sky regions can the polarized loop emission be identified?
\\

\textbf{Methods}: Based on a well known result concerning the magnetic field
alignment in supernova explosions, a theoretical expectation is established
that the loop polarizations should be predominantly E-mode. In particular, the
expected polarization angles of Loop I are compared with those from the real
microwave band data of WMAP and Planck.\\

\textbf{Results and conclusions}: The comparison between model and data shows
remarkable consistency between the data and our expectations at all bands and
for a large area of the sky. This result suggests that the polarized emission
of Galactic Loop I is a major polarized component in all microwave bands from
23 to 353 GHz, and a considerable part of the polarized foreground likely
originates from a local bubble associated with Loop I, instead of the far more
distant Galactic emission. This result also provides a possible way to explain
the reported E-to-B excess~\citep{2016A&A...586A.133P} by contribution of the
loops. Finally, this work may also provide the first geometrical evidence that
the Earth was hit by a supernova explosion.\\

}

\mktt

\section{Introduction}\label{sec:intro}

Galactic Loop I is a large circular structure on the north Galactic sky whose
brightest part is also referred to as the north polar spur
(NPS)~\citep{Berkhuijsen1971, Salter1983}. It shines from the radio band to
the $\gamma$-ray band~\citep{1981A&A...100..209H, Haslam1981}, including
microwaves~\citep{2013ApJS..208...20B, 2016A&A...594A...1P}, and possibly even
affects the cosmic rays~\citep{bhat_acceleration_1985}. The origin of this
structure is suggested to be an old supernova~\citep{Berkhuijsen1971,
Salter1983, Wolleben2007} that created its own local bubble which happens to
be in close contact with the Orion local bubble~\citep{1995A&A...294L..25E,
2006A&A...452L...1B}. The brightest part of Loop I is a few tens of degrees in
width, and is about $60\degree$ away from its center, whose sky direction is
around $(l,b) = (329\degree, 17.5\degree)$~\citep{Berkhuijsen1971, Mertsch}.
The distance of the old supernova is not well determined, but could be of the
order of $10^2$ pc~\citep{Mertsch}. Estimating from its roughly $60\degree$
angular radius, the wavefront of the explosion must have already traveled at
least half the distance between its point of origin and Earth.

Although it is quite natural to imagine that the Earth could have been hit by
a supernova explosion, to date, there is only indirect evidence provided by
statistical expectation~\citep{WHITTEN1976, 1977Natur.265..318C}; oceanic
traces of $^{60}$Fe~\citep{2004PhRvL..93q1103K},
$^{44}$Ti~\citep{1999NewA....4..419F, 2000APh....14....1B}, and other
isotopes~\citep{2005ApJ...621..902F}; and a combined
estimation~\citep{2002PhRvL..88h1101B}. In principle, much more direct
evidence could be obtained by geometric considerations: before the explosion
hits the Earth, observers facing the supernova can only see its signal from
the front side. Therefore, if one sees the supernova signal from both front
and back sides, then the Earth must have been hit by the supernova explosion.
The most difficult part of this idea is how to associate a signal coming from
the back side with a supernova remnant lying in the front side. Fortunately,
in this work it is shown that this problem can be solved by comparing the
polarization angles measured in the microwave bands with the assumption of a
minimal Loop I model.

As briefly introduced in Appendix~\ref{app:EB in real space}, the polarized
signals can be decomposed into the E and B modes. It was suggested
by~\cite{2018arXiv180410382L} that for a better foreground removal, such EB
decomposition should be done in form of
\begin{equation}\label{equ:basic decomposition}
(Q,U)\equiv(Q_E,U_E)+(Q_B,U_B).
\end{equation}
If the $(Q_E,U_E)$ and $(Q_B,U_B)$ signals are found to be associated with
different physical mechanisms, then such decomposition is a natural choice.
This is confirmed in this work by showing that the loop polarizations are
predominantly E-mode.

The paper is organized as follows: Section~\ref{sec:loops and E-mode} explains
why the loop polarizations are predominantly E-mode, which is also observable
in real data. Section~\ref{sec:minimal model} proposes a simplified model for
the Loop I polarization angles that contains \emph{no free parameters}, and
which is found to be consistent with the real foregrounds at 99.999$\%$
confidence level. This strongly supports that the polarized microwave
foregrounds are largely from Loop I, as well as that the Earth has been hit by
the Loop I supernova explosion. Finally, the results are discussed in
Section~\ref{sec:discussion} and conclusions are given in
Section~\ref{sec:conclusion}.

\section{Loops and E-mode}\label{sec:loops and E-mode}

\subsection{Why the loops are predominantly E-mode}\label{sub:why loops are E-mode}

This work starts from a simple model of the loop polarizations, which allows
several trivial parameters (position, size, radial profile, intensity, etc.)
but there is \emph{only one} major assumption: signals from the loop are
polarized along the radial directions of the supernova. This assumption is
based on the conclusion that the shell magnetic fields are along the tangent
directions~\citep{1962MNRAS.124..125V}, which was confirmed
by~\citep{1968ApJ...154..807W, 1987AuJPh..40..771M}, and recently reviewed and
developed by~\citep{2015A&ARv..23....3D, 2016MNRAS.456.2343P}. We also note
that this major assumption is not a new one: application and verification of
this assumption on the microwave maps can also be found in
\cite{2015MNRAS.452..656V}, for example.

\begin{figure}
  \centering
  \includegraphics[width=0.23\textwidth]{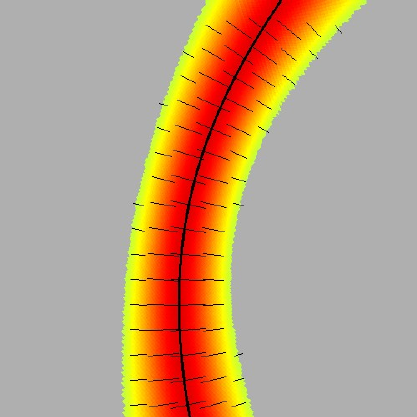}
  \includegraphics[width=0.23\textwidth]{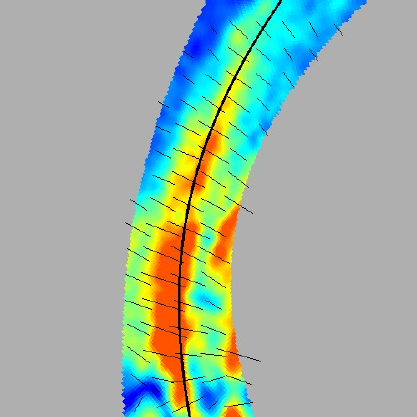}

  \includegraphics[width=0.48\textwidth]{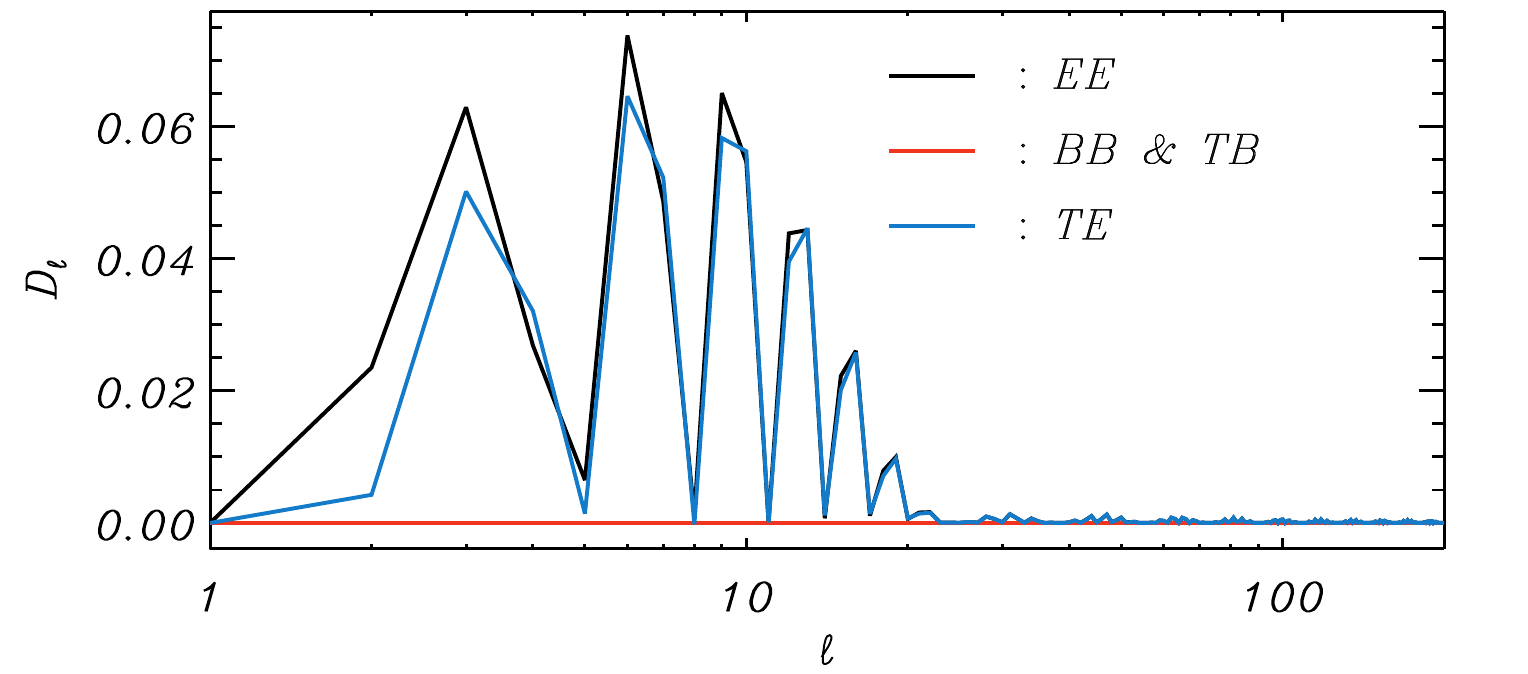}
  \caption{\emph{Upper left}: Example of the expected loop polarization
  directions (by small black lines). \emph{Upper right}: Polarization
  directions of the WMAP K-band in the Loop I region, with Loop I marked by a
  black circle. The color scales of the upper panels are for the polarized
  intensity, which is 0$\sim$100 $\mu$K for the K-band and -1$\sim$1.2 for the
  model (here the absolute amplitude of polarization intensity is meaningless
  for the model, and therefore such an unphysical range is chosen to maximize
  the visibility of the thin lines). \emph{Lower}: the angular power spectrum
  of the model shown in upper-left, calculated without a mask. We note that
  the polarization directions in the model are strictly along the normal
  vectors, and some small misalignments are only visual effects due to
  pixelization.}
  \label{fig:loop model example}
\end{figure}

More discussions  supporting this major assumption can be found in
Appendix~\ref{app:obs illustration}, but the assumption itself is pure
geometry and is easy to apply. A typical model for Loop I polarization is
therefore generated, as follows: First I take a Gaussian radial profile for
the polarization intensity, which is maximized and equal to 1 at $58\degree$
to the center of Loop I ($58\degree$ is the angular radius of the brightest
part of Loop I), and with a $20\degree$ FWHM (the FWHM parameter will affect
the shape of the angular power spectrum, but here we mainly pay attention to
``zero or not'', and therefore this parameter is not important). Then the
polarization angles are set to along the radial directions of Loop I. The
result of this model is shown in Figure~\ref{fig:loop model example} and
compared with the real Loop I polarization angles at the WMAP K-band (22.8
GHz) with a $20\degree$ wide mask to emphasis the NPS region. However, we note
that the model is actually full sky, with the main power being localized
around Loop I. This is very convenient in avoiding the leakage from E-mode to
B-mode, because such leakage does not exist for a full sky map. The full sky
angular power spectra of the model is calculated and shown in the lower panel
of Figure~\ref{fig:loop model example}.

The excellent agreement between the polarization angles of the model and data
shown in the upper panels of Figure~\ref{fig:loop model example} fully
supports the above major assumption. Subsequently, by connecting
Figure~\ref{fig:loop model example} with the geometrical interpretation of the
E and B modes in real space (see Appendix~\ref{app:EB in real space}), one can
see that this major assumption can produce only E-mode polarizations. This is
confirmed by the angular power spectra of the model in Figure~\ref{fig:loop
model example}, which is calculated as described above. The angular power
spectra have positive EE \& TE spectra, but these are zero for BB \& TB. We
note that this important property depends only on the major
assumption\footnote{\emph{All} other parameters of this basic model are
trivial and can be eliminated to make a minimal model, which is done in
Section~\ref{sec:minimal model}. Meanwhile, the major assumption certainly
allows small deviations/fluctuations, which is briefly discussed in
Section~\ref{sec:discussion}.}.

Therefore, polarized loop signals will form an E-mode foreground family, whose
presence will certainly provide net EE and TE excess. This can be naturally
associated with the E-mode excess reported by~\cite{2016A&A...586A.133P} in
terms of an E-to-B ratio roughly equal to 2. This is discussed in
Section~\ref{sub:E-mode excess}.

\subsection{E-mode loops/arches in real data}\label{sub:E family and loops}

\begin{figure*}[!htb]
  \centering
  \includegraphics[width=0.32\textwidth]{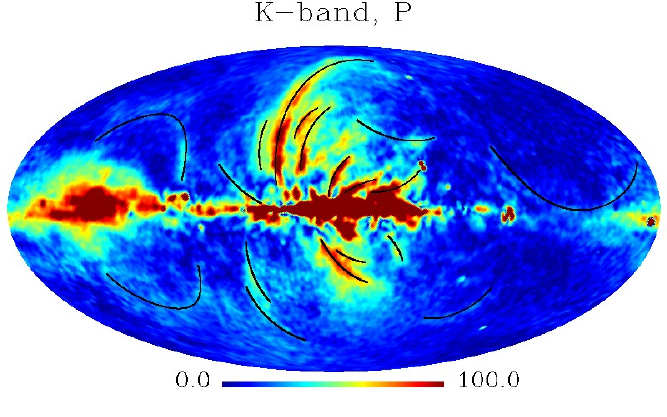}
  \includegraphics[width=0.32\textwidth]{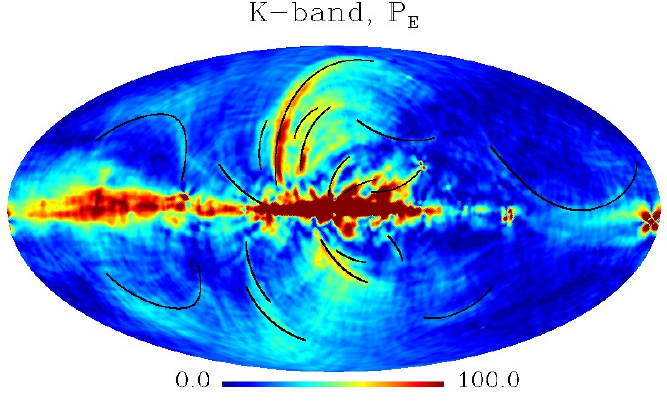}
  \includegraphics[width=0.32\textwidth]{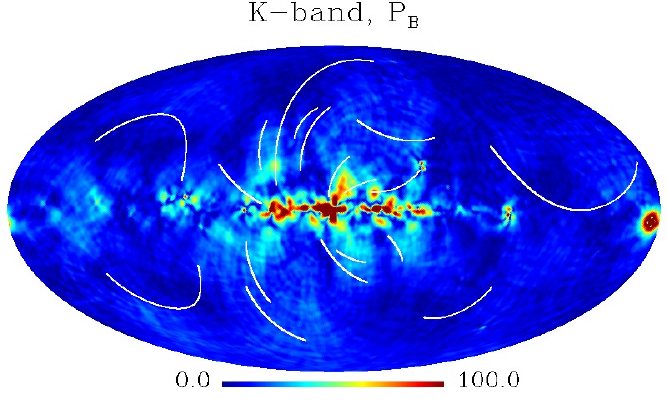}

  \includegraphics[width=0.32\textwidth]{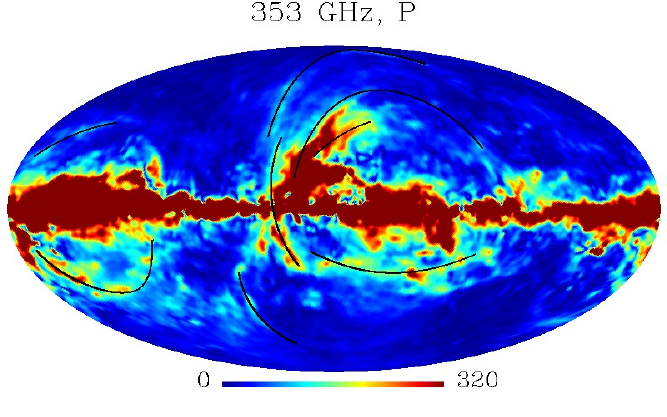}
  \includegraphics[width=0.32\textwidth]{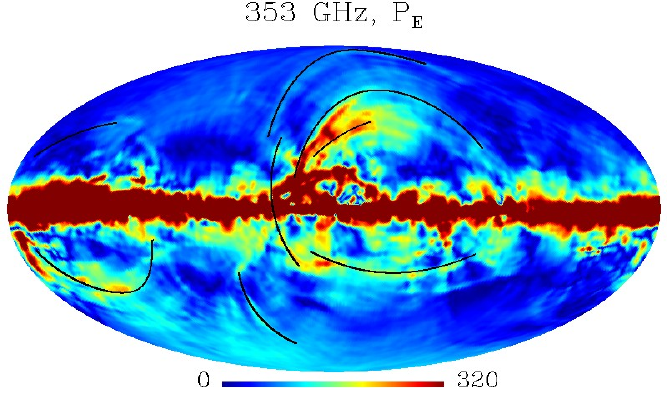}
  \includegraphics[width=0.32\textwidth]{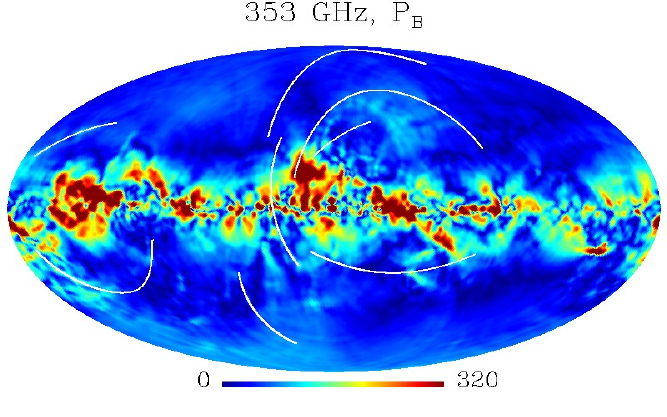}
  \caption{ The polarization intensity maps and their EB decompositions at the
  K-band (upper) and 353 GHz (lower). From left to right: total polarized
  intensity $P$, $P_E$, and $P_B$. The loop-like structures are marked by
  black or white lines depending on visibility. They are apparently visible in
  E-mode (middle) but disappear in B-mode (right).}
  \label{fig:P Pe and Pb}
\end{figure*}

To directly investigate the E-mode loops in real data, the WMAP K-band and
Planck 353 GHz full sky polarization maps are decomposed using
Equation~\ref{equ:basic decomposition}, and the E and B mode polarization
intensities calculated as $P_E=\sqrt{Q_E^2+U_E^2}$ and
$P_B=\sqrt{Q_B^2+U_B^2}$,  respectively. The results of such decomposition are
shown in Figure~\ref{fig:P Pe and Pb}. Apparently, after decomposition, all
loop structures\footnote{Several arches are placed along the loops in
Figure~\ref{fig:P Pe and Pb} similar to~\cite{2016A&A...594A..25P,
2015MNRAS.452..656V}. Especially for the K-band, we refer
to~\cite{2015MNRAS.452..656V} and make the arches consistent with their Figure
2, except that a few of the arches are split into two for better match.} are
visible in $P_E$ but disappear in $P_B$, which strongly supports the
conclusion that loop polarizations are an E-mode foreground family. This was
also tested by~\cite{2018arXiv180410382L}.

\begin{figure*}[!htb]
  \centering
  \includegraphics[width=0.32\textwidth]{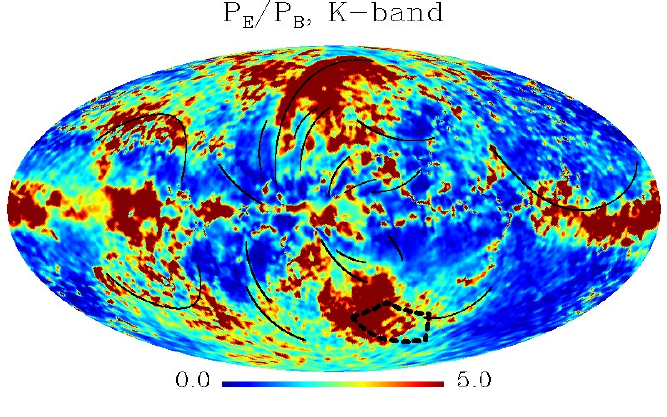}
  \includegraphics[width=0.32\textwidth]{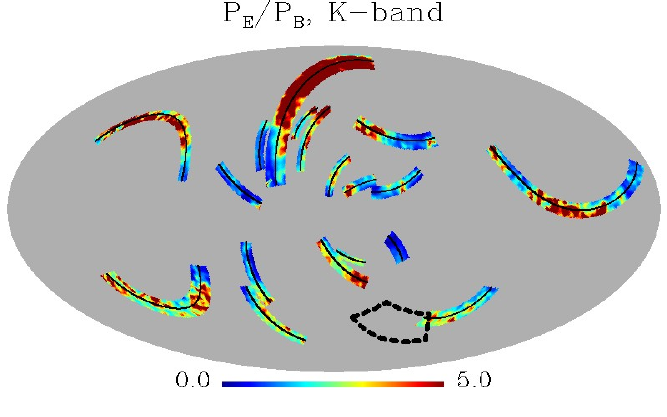}

  \includegraphics[width=0.32\textwidth]{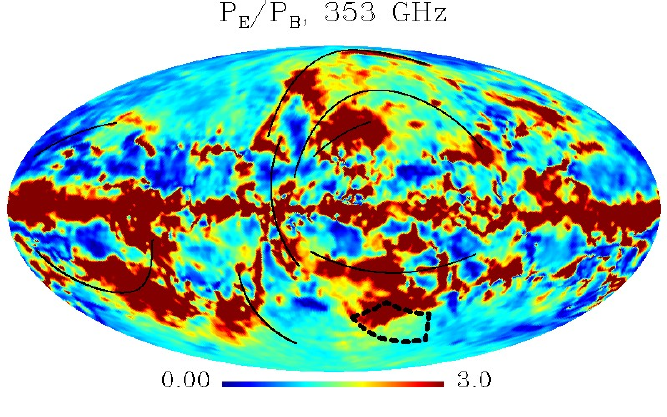}
  \includegraphics[width=0.32\textwidth]{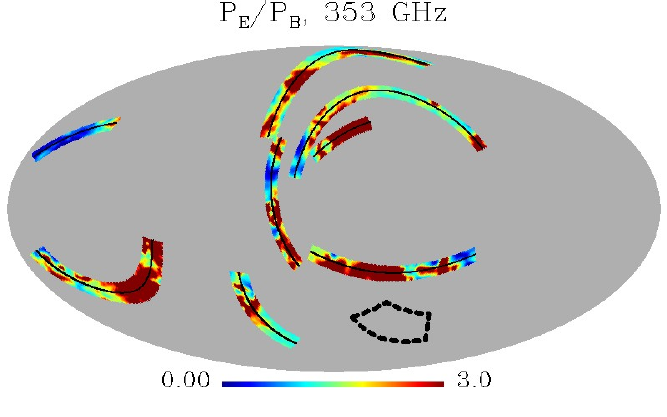}
  \caption{ The ratio $\rho=P_E/P_B$ in K-band (upper) and 353 GHz (lower),
  both full sky and around the same arches as marked in Figure~\ref{fig:P Pe
  and Pb}. The outline of the BICEP2 zone is marked in thick dash
  line.}
  \label{fig:r K and 353}
\end{figure*}

The ratio between the E and B mode polarization intensities is calculated as
$\rho=P_E/P_B$ and presented in Figure~\ref{fig:r K and 353}, especially for
the arch regions. The median and mean values of $\rho=P_E/P_B$ are compared
for inside and outside the arch regions, and the results are listed in
Table~\ref{tab:pe/pb}. For all cases, $\rho$ are apparently higher for the
inside regions, which further supports the argument that loop polarizations
are mainly E-mode. A simple test is then done to show the significance of the
values in Table~\ref{tab:pe/pb}: For each band, the corresponding arch mask is
rotated to 192 evenly distributed directions (corresponding to $N_{side}=4$),
and for each direction, the new mean value is calculated inside the new mask.
For the K-band, none of the 192 new directions give a higher mean value of
$\rho$ than the unrotated one (5.4), therefore the confidence level is at
least $99.5\%$. Similarly, for 353 GHz the confidence level is $99\%$.

\begin{table}
 \caption{The median and mean values of $\rho$ for the region inside/outside
  the arch regions shown in Figure~\ref{fig:r K and 353}. }
 \centering
 \begin{tabular}{|c|c|c|} \hline
          & Median           & Mean \\
          & inside (outside) & inside (outside) \\ \hline
  K-band  & 2.6 (1.9)        & 5.4 (2.8) \\ \hline
  353 GHz & 1.9 (1.3)        & 3.0 (2.1) \\ \hline
 \end{tabular}
 \label{tab:pe/pb}
\end{table}

\subsection{Loops and E-mode excess}\label{sub:E-mode excess}

It was reported by the~\cite{2016A&A...586A.133P} that there is an excess of
the E-mode foreground in the 353 GHz band with an EE-to-BB ratio of
approximately two, and possible explanations were proposed based on the MHD
properties; see, for example, \cite{2017ApJ...839...91C},
\cite{2017MNRAS.472L..10K} and \cite{ 2017arXiv171111108K}. With the proposals
from Sections~\ref{sub:why loops are E-mode}--\ref{sub:E family and loops}, a
parallel direction is opened, in which the E-mode excess can be naturally
explained by the existence of the loops. Detailed works in this direction will
follow, and a tentative estimation is that, due to the reported E-to-B ratio,
the loops can contribute as much as the ordinary diffuse Galactic foreground
emission, which is expected to have similar E and B mode spectra.

We also note that, for the loop polarizations, the shape of their E-mode angular
power spectra is determined by the choice of the trivial parameters. Since the
real sky can possibly contain many loop-like
structures~\citep{2004A&A...418..131K, 2007A&A...463.1227K, Mertsch,
2015MNRAS.452..656V}, it is quite possible to fit the observed polarized
foreground spectrum by a family of loops, which is an interesting direction
for future work.

\section{Model without free parameters}\label{sec:minimal model}

An important adjustment of the basic model discussed in Section~\ref{sub:why
loops are E-mode} is to eliminate \emph{all} free parameters and fix the
model. This has the great advantage that no fitting is needed at all, and
the model is therefore completely independent from the data.

This is done as follows: 
\begin{enumerate}
\item Only the polarization angles are considered, so all parameters about
        amplitude become irrelevant.
\item The Loop I emission is regarded as coming from the associated
  bubble\footnote{In this work, the angular radius of the local bubble is not
  necessarily the same as the angular radius of Loop I, which is
  $\sim$$60\degree$.} whose radius can be bigger than the distance to its
  center, such that the emission can cover the full sky.
\item Finally, with the major assumption mentioned at the beginning of
        Section~\ref{sub:why loops are E-mode} and the already known $(l,b)$
        coordinates of the Loop I central supernova remnant, all polarization
        angles can be calculated without any other parameters or assumptions.
\end{enumerate}

Following the above procedures, one obtains a \textbf{full sky} pattern of the
Loop I polarization angles. We note that, although the emissions from Loop I
were not regarded as covering the full sky before, people have already
discussed the possibility that the signal from Loop I is extended beyond the
NPS region; for example, \cite{2016A&A...594A..25P} and
\cite{2015MNRAS.452..656V}.

\subsection{Visual inspection}\label{sub:result and visual}

\begin{figure*}[!htb]
  \centering
  \includegraphics[width=0.32\textwidth]{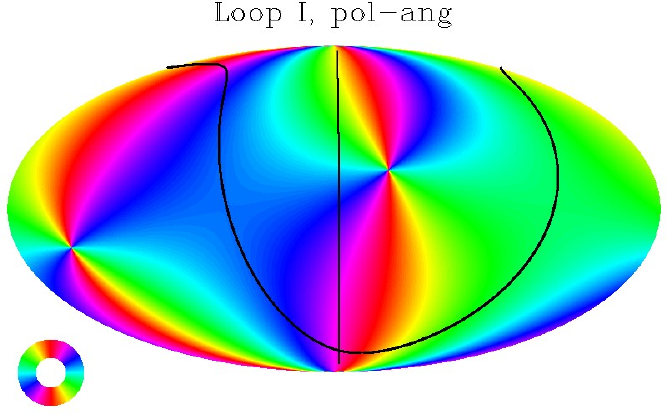}
  \includegraphics[width=0.32\textwidth]{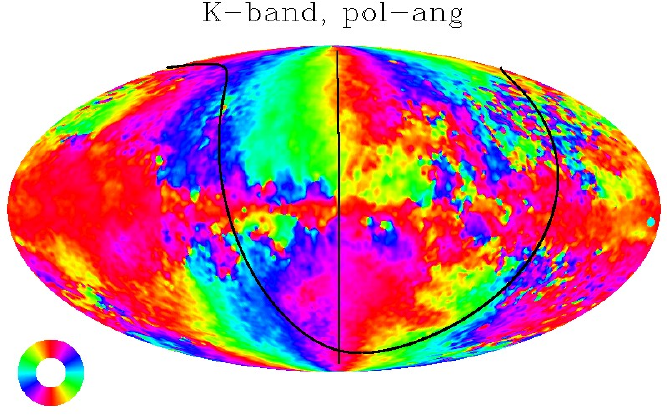}
  \includegraphics[width=0.32\textwidth]{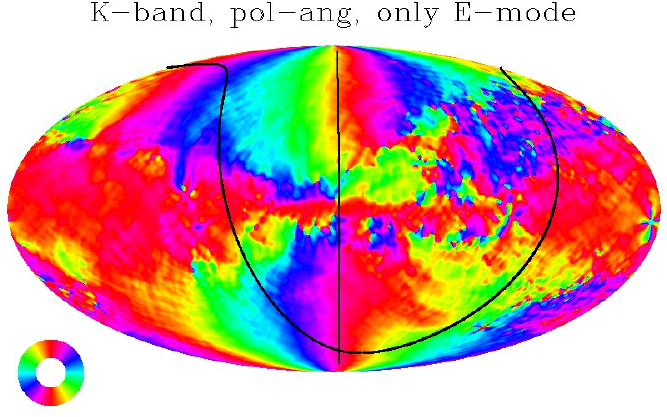}
  \caption{ \emph{Left}: Polarization angles calculated from the minimal
  model, with no free parameters and no fitting. \emph{Middle}: the real data (WMAP
  K-band). \emph{Right}: real data and only E-mode. The color-to-angle mapping
  is given by the low-left color disk, and the black circle is the critical
  range that is $90\degree$ from the center of Loop I.}
  \label{fig: compare model to data}
\end{figure*}

A completely determined model means no need for fitting, and therefore the full sky
polarization angles can be calculated straightforwardly and compared with the
polarization angles measured at the WMAP K-band ($2\degree$ smoothing), as
shown in Figure~\ref{fig: compare model to data}. By simple visual
inspection, one can easily see qualitative similarity between the model and
data for the greater part of the sky: In the north sky, starting from the central
black line, the color on the left-hand side starts from green and goes
\emph{anti-clockwise} along the color disc\footnote{See the color disk in the
lower left-hand corner of each panel in Figure~\ref{fig: compare model to data} for the
rotational color-to-angle mapping.} until green again; while in the right hand
side (still north) the color starts from yellow and goes \emph{clockwise} to
green. Similarly, in the south sky the color variation left of the central
black line is \emph{clockwise} from purple to blue, and for the right-hand
side it is \emph{anticlockwise} from purple to blue. All these color rotations
are the same for both the data and the Loop I model, which indicates that the
polarized microwave emission is associated with Loop I for a large area of the
sky.

We note that the model-to-data comparison can be made more fair by considering
only the E-mode of the WMAP K-band data. Doing so, one does see
better data-to-model consistency, which is expected, especially in the lower-right corner of the
map, as also presented in Figure~\ref{fig: compare model to data}.

\subsection{Significance estimation}\label{sub:significance}

The visual inspection in Section~\ref{sub:result and visual} is good for
understanding, but is not quantitative. For example, the red and yellow colors
are visually very different, but the corresponding angles can be close to each
other. Therefore, the consistency between data and model is quantitatively
tested below using the mean angle difference (MAD) $\<\delta_\theta\>$ defined
below.

Allowing the polarization angle difference to be $\delta_\theta'=\theta_1-\theta_2$,
by the definition of polarization angles, $\delta_\theta'$ is identical
to $\delta_\theta'\pm180\degree$. Moreover, if one disregards the sign,
then $\delta_\theta'$ is also identical to $-\delta_\theta'$. To reflect these
symmetries, in this work, the polarization angle difference is defined as
\begin{eqnarray}\label{equ:delta theta}
\delta_\theta &=& |\arcsin(\sin(\delta_\theta'))| \\ \nonumber
&\equiv& 90\degree - |90\degree-\arccos[\cos(\delta_\theta')]|,
\end{eqnarray}
where $\delta_\theta$ always lies in the range $0\degree$--$90\degree$, and all
four angles $\pm\delta_\theta'$ and $\pm(\delta_\theta'-180\degree)$ are
regarded as equivalent. Subsequently, the MAD $\<\delta_\theta\>$ is defined as
\begin{eqnarray}\label{equ:mad}
\<\delta_\theta\> = {\rm{arctan2}}\left(\sum\sin(\delta_\theta),\sum\cos(\delta_\theta)\right),
\end{eqnarray}
where the ${\rm{arctan2}}$ function is a variant of the $\arctan$ function
that takes two parameters to return a result in the range $[0,2\pi]$. With the
above definition, the similarity between two sets of angles can be roughly
evaluated by $\cos(\<\delta_\theta\>)$, where $\cos(\<\delta_\theta\>)=1$
indicates that the two sets of angles are identical. For two independent maps, the
expectation of $\<\delta_\theta\>$ should be centered around $45\degree$, and
due to the central limit theorem, for data sets with more degrees of freedom
(such as more pixels), the distribution of $\<\delta_\theta\>$ is more narrow
and Gaussian.

\begin{figure*}[!htb]
  \centering
  \includegraphics[width=0.32\textwidth]{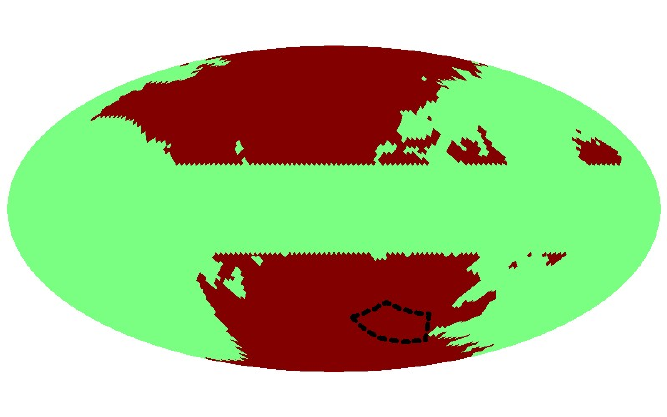}
  \includegraphics[width=0.32\textwidth]{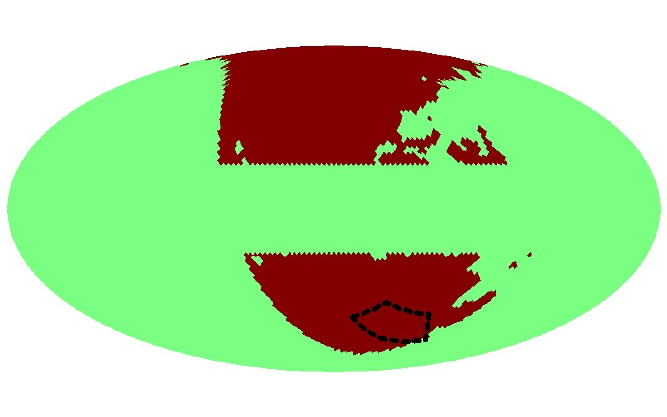}
  \includegraphics[width=0.32\textwidth]{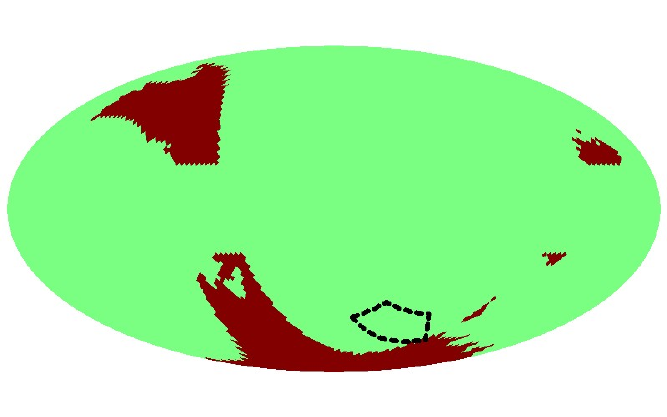}
  \caption{The regions used to estimate the significance of the model-to-data
  association, where red means usable region and green means masked region.
  \emph{Left}: the region used in Section~\ref{sub:significance}. \emph{Middle
  \& right}: the inside and outside regions used in
  Section~\ref{sub:inference}, which are $0\degree$--$90\degree$ and
  $90\degree$--120$\degree$ to the center of Loop I, respectively. The BICEP2
  region is also marked in each panel.}
  \label{fig:regions}
\end{figure*}

The sky region at low Galactic latitudes is dominated by the strong emission
from the Galactic plane, and therefore a ring mask is used to exclude
$|b|\le20\degree$. Meanwhile, the area with very low polarized signal is
dominated by noise, which is uncorrelated with any real signal, and therefore a
fraction $\rho=25\%$ of the sky\footnote{It was confirmed that the result does
not change significantly for $\rho=20\%$--$30\%$, so in the following calculations
$\rho=25\%$ is adopted.} with the lowest $P_E$ is also excluded. The loop
signal is expected to decay with increasing radius to the center, and so only
the region less than $120\degree$ from the center of Loop I is used. The
combined mask is shown in the left panel of Figure~\ref{fig:regions}.

\begin{figure}[!htb]
  \centering
  \includegraphics[width=0.48\textwidth]{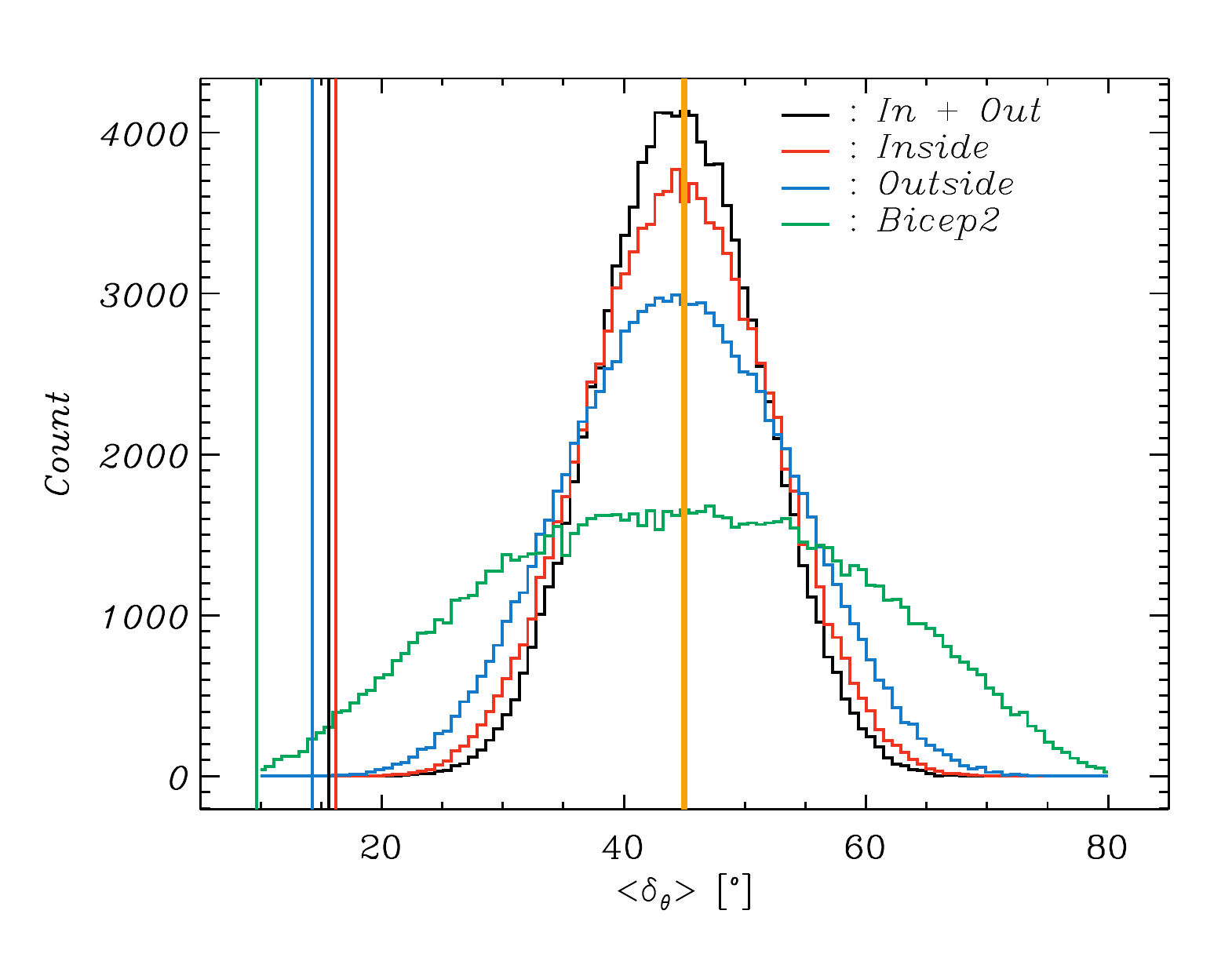}
  \caption{Histograms of $10^5$ mean angle differences $\<\delta_\theta\>$
  between the minimal model and simulations, on which results from the real
  data are marked by thin vertical lines, and the $45\degree$ expectation of
  $\<\delta_\theta\>$ is marked by the central vertical line.}
  \label{fig:hist mad}
\end{figure}

The MAD between the real K-band map and the minimal model is first calculated
with the mask shown in the left panel of Figure~\ref{fig:regions}, which is
$\<\delta_\theta\>=15.6\degree$, with $\cos(\<\delta_\theta\>)=0.96$, very
close to 1. Realistic simulations are then run: First the input
$(Q,U)$ Stokes parameter map is converted into $\alpha_{lm}^{EB}$ using
HEALPix~\citep{2005ApJ...622..759G}. Subsequently, the phases for
$\alpha_{lm}^E$ and $\alpha_{lm}^B$are randomized before inverse transforming to
real space using HEALPix to get a simulated map. Since the EE and BB spectra
are unaffected by the phases, such an operation completely changes the
morphology of the input map without changing its E and B spectra. We note that
the B-mode is set to zero in the simulations because the model is also pure
E-mode. Using $10^5$ simulations generated like this, I find none that yield
a MAD below $20\degree$ for the region in use, as illustrated by the
histograms of all simulations in Figure~\ref{fig:hist mad}. According to this
test, the E-mode of the K-band foreground map gives polarization angles that
are consistent with those calculated in our minimal model at a $99.999\%$
confidence level\footnote{I note that the real foreground emissions are
non-Gaussian and inhomogeneous, and therefore it is very difficult to make a random
simulation that can fully reproduce all physical and statistical properties of
the polarized foreground. The simulations here faithfully reproduce the
angular spectrum of the foreground, but are still imperfect in reproducing the
characteristic phases of $\alpha_{lm}^E$ and $\alpha_{lm}^B$ that represent
the non-Gaussian, inhomogeneous properties of the foreground.}

\subsection{Critical range and the BICEP2 region}\label{sub:inference} 

Now I discuss how to
determine whether the supernova has hit the Earth or not. The division between
the front and back sides is made at $90\degree$ from the loop center, which is
marked by a black circle in Figure~\ref{fig: compare model to data}. If the
similarity between the minimal model and the data is significant for both
sides of the black circle, then it is suggested that the Earth has been hit by
the Loop I supernova explosion.

For this purpose, the region shown in the left panel of
Figure~\ref{fig:regions} is divided into two sub-regions and shown in the same
figure as the middle and right panels: the middle panel shows the inside
region (front side, from 0$\degree$ to $90\degree$), and the right panel shows
the outside region (back side, from $90\degree$ to $120\degree$). The
resulting MAD for the inside region is $\<\delta_\theta\>=16.2\degree$, with
$\cos(\<\delta_\theta\>)=0.96$; and for the outside region is
$\<\delta_\theta\>=14.3\degree$, with $\cos(\<\delta_\theta\>)=0.97$. Using
$10^5$ simulations generated as described in Section~\ref{sub:significance}
for the two regions respectively, I find the values of both regions to be
significant at a $99.999\%$ confidence level, which are also shown in
Figure~\ref{fig:hist mad} by red and blue lines. Therefore, both the inside and
outside regions are suggested to be associated with Loop I, which means we are
likely sitting inside the bubble of Loop I.

Meanwhile, the above calculation is also done for the BICEP2 region shown in
Figure~\ref{fig:regions} and the result is included in Figure~\ref{fig:hist
mad}, which has an exceptionally low value of $\<\delta_\theta\>=9.7\degree$
and $\cos(\<\delta_\theta\>)=0.99$. Due to the smaller size of the BICEP2
region, the distribution of $\<\delta_\theta\>$ is much wider and is
apparently non-Gaussian. Using $10^5$ simulations, I find the association
between Loop I and the BICEP2 region is still confirmed at $99.96\%$.

\subsection{Other frequencies and CMB}\label{sub:minimal model for other frequencies}

The analyses presented in Sections~\ref{sub:significance}--\ref{sub:inference}
are also done for all the WMAP and Planck frequency bands that contain
polarization data, including the WMAP K, Ka, Q, V, W; and the Planck 30, 44,
70, 100, 143, 217, 353 GHz. The Planck SMICA CMB map with polarization is
subtracted from each of them to roughly remove the CMB\footnote{Due to its low
amplitude, all results in this work are nearly the same with/without
subtracting the CMB.}, and the Planck LFI bandpass mismatch correction is
applied to 30, 44, and 70 GHz bands. A complete list of $\<\delta_\theta\>$ is
presented as Table~\ref{tab:mad}, in which one can see that all of them give
apparently lower $\<\delta_\theta\>$ than the $45\degree$ expectation, where
the minimal confidence level is no less than $99\%$ for each band. This means
all WMAP and Planck frequency bands are significantly contaminated by the Loop
I polarized emissions. If all bands are regarded as independent, then the
combination of them gives a surprisingly high confidence level.

I also note that the Planck 30 and 44 GHz bands (marked in blue in the table)
give apparently higher $\<\delta_\theta\>$ than their neighbors, especially the 30
GHz band. This is most likely due to the bandpass mismatch
leakage~\citep{2016A&A...594A...2P, 2016A&A...594A...3P} that remains even
after correction~\citep{2018arXiv180101226W}. A similar abrupt
$\<\delta_\theta\>$ value exists for the Planck 100 GHz band in the BICEP2
region, which is also marked in blue. On the other hand, the case for the
Planck 70 GHz band is slightly unclear: it is also contaminated by the
bandpass mismatch leakage, but the amplitude of contamination is regarded as
less than 30 and 44 GHz. One can still see from Table~\ref{tab:mad} that 70
GHz has moderately higher $\<\delta_\theta\>$ than its neighbors, which could be
due to either the remaining bandpass mismatch leakage, or relatively lower
polarized foreground at 70 GHz. The latter could be good news for detection of
the primordial B-mode; however, since a full consideration of the systematics
is very complicated, the above discussions are only suggestive. An updated
study will be possible when the future Planck data release becomes available, which
may come with lower systematic errors.

\begin{table*}[!htb]
 \caption{List of $\<\delta_\theta\>$ in degrees between the minimal Loop I
 model and 23-353 GHz bands. $\<\delta_\theta\>\ll45\degree$ indicates
 apparent model-data correlation. Some cells are suspicious because they are
 apparently higher than the neighboring bands; they are marked in blue.
 All unmarked values are apparently lower than the $45\degree$ expectation.
 The BICEP2 region is particularly well correlated with Loop I for all
 frequency bands except for the suspicious values. }
 \centering
 \begin{tabular}{|c|c|c|c|c|c|c|c|c|c|c|c|c|} \hline
  \rowcolor{gray!40!}
  Band      &      K &     30                 &     Ka &      Q &     44                    &      V &   70   &      W &    100 &    143 &    217 &    353 \\ \hline
  \rowcolor{gray!40!}
  $\nu$ (GHz)         
            & 22.8   &   28.4                 &   33.0 &   40.7 &   44.1                    &   60.8 &   70.4 &   93.5 &    100 &    143 &    217 &    353 \\ \hline 
  In \& Out & 15.6 &  \cellcolor{blue!40}26.9 &   16.5 &   16.7 &   \cellcolor{blue!40}28.3 &   29.6 &   31.7 &   22.0 &   17.1 &   20.0 &   18.6 &   19.7 \\ \hline 
  Inside    & 16.2 &  \cellcolor{blue!40}25.0 &   16.6 &   16.4 &   \cellcolor{blue!40}26.1 &   27.4 &   28.7 &   22.2 &   17.3 &   19.8 &   18.3 &   19.8 \\ \hline 
  Outside   & 14.3 &  \cellcolor{blue!40}31.1 &   16.2 &   17.3 &   \cellcolor{blue!40}32.5 &   34.5 &   36.8 &   21.7 &   16.8 &   20.6 &   19.3 &   19.5 \\ \hline 
  BICEP2    & 9.7  &  \cellcolor{blue!40}53.5 &    9.0 &    6.8 &   \cellcolor{blue!40}27.2 &   21.8 &   30.2 &   19.5 &   \cellcolor{blue!40}35.5 &    5.3 &    4.1 &    7.7 \\ \hline 

 \end{tabular}
 \label{tab:mad}
\end{table*}

The value of $\<\delta_\theta\>$ is also calculated between the minimal model and the
Planck SMICA CMB map. In this case, for the three masks shown in
Figure~\ref{fig:regions}, $\<\delta_\theta\>$ are 41.4$\degree$, 43.1$\degree$
and 38.5$\degree$, respectively. Interestingly, these values are much closer to
$45\degree$, but are still systematically lower, which indicates a possible
residual loop contamination even in the final CMB polarization product.

\section{Discussion}\label{sec:discussion}

It was pointed out by \cite{liu_fingerprints_2014} and
\cite{von_hausegger_footprints_2016} that the Galactic Loop I may leave a
trace on the final CMB intensity map, which is partly verified in this work
for the E-mode. This provides a good reason to follow the suggestions
by~\cite{2018arXiv180410382L} to adopt the decomposition in
Equation~\ref{equ:basic decomposition}, which may help to improve the
estimation of the CMB B-mode.

For an incomplete sky coverage (which is the case for all individual
ground missions), the above decomposition is inevitably affected by the E-to-B
and B-to-E leakage. Although there are already many methods to prevent such
leakages~\citep{2006PhRvD..74h3002S, 2010A&A...519A.104K, PhysRevD.82.023001,
2017PhRvD..96d3523B, 2018arXiv180105358K}, they are mainly designed for
Gaussian and homogeneous CMB signals, and are therefore problematic for
non-Gaussian, inhomogeneous signals such as diffuse Galactic foregrounds.
Therefore, a large (or even full) sky coverage -- either by combining various
ground missions or from a space mission~\citep{2012SPIE.8442E..19H,
1475-7516-2018-04-018} -- is apparently preferred for detection of primordial
gravitational waves.

If the supernova explosion is spherically symmetric, then by our major
assumption, the loop emission is $100\%$ E-mode. However, in reality, the
supernova explosion could be asymmetric, and therefore there could be residual B-mode
emission from the loops. For example, a multiple supernova explosion scenario
was studied by~\citep{2017MNRAS.468.2757V}, in which the overall shape of the
shell is naturally asymmetric. Also, in Figure~\ref{fig:P Pe and Pb}, although
much fainter, one can still see some suspicious loop-like structures in the
$P_B$ map, which might be this kind of residual. Meanwhile, another source of
the B-mode from loops due to projection is also discussed in
Appendix~\ref{app:obs illustration}. 

Recent work~\citep{2018arXiv180410382L} has confirmed that for both the
E-mode and the B-mode families in the BICEP2 zone, the polarization angles are
almost the same from 217 to 353 GHz; meanwhile, Table~\ref{tab:mad} tells us that
in the BICEP2 zone, the polarization angle is tightly related to Loop I. These
two facts highlight the possibility that the B-component from Loop I (if any)
may also affect the high Galactic latitudes, such as the BICEP2 area.

It should also be noted that the LSA model~\citep{2007ApJS..170..335P} for the
Galactic magnetic field can also give results that are  consistent with the
WMAP K-band polarization angles, and can explain the typical foreground
polarization fraction. The disadvantage of this model is that it requires the
fitting of several free parameters regarding the Galactic spiral structure and
the high-energy electron distribution, whereas the model in
Section~\ref{sec:minimal model} has no free parameters and requires no
fitting, making it the preferable option. Moreover, to avoid circular
argument, a model based on fitting can only provide indications of the general
trends of the data, and cannot be used for further purposes, such as
explaining the E-mode excess -- which can be done easily and naturally using a
study such as the one presented here. In reality, however, the line-of-sight
integration for the polarized signal consists of both local and remote parts,
and therefore reality is more likely to be better represented by a combination
of this work and the LSA model.

The polarized signal can change its polarization direction due to integration
along the line-of-sight (LOS), and such integration can easily decrease the
total polarization intensity, which is called depolarization. Such
depolarization is considered both by~\cite{2007ApJS..170..335P} and in
Appendix~\ref{app:obs illustration}, as well as in studies of three-dimensional foreground analysis~\citep{2008A&A...477..573S,
2011A&A...526A.145F, 2015ApJ...810...25G, 2017arXiv170604162M}. All these works
depend on our knowledge of the Galactic and local magnetic field, which remains far from perfect. Therefore, a three-dimensional
analysis for polarized foreground still requires better constraints.


\section{Conclusion}\label{sec:conclusion}
The main conclusions of this work are listed below:
\begin{enumerate}
\item The supernova explosions can produce predominantly E-mode foreground
  (Section~\ref{sub:why loops are E-mode}).
\item The E-mode loops provide a new way to explain the E-to-B excess
  phenomenon (Section~\ref{sub:E-mode excess}).
\item A large part of the polarized foreground is likely coming from a local
  bubble associated with Loop I, which suggests that the Earth was hit by a
  supernova explosion (Section~\ref{sub:inference}).\label{item:conclusion hit
  earth}
\item The E-mode foreground from Loop I is identified at all WMAP and Planck
  frequency bands and for a large area of the sky, including the high Galactic
  latitudes (Section~\ref{sub:minimal model for other frequencies}). However,
  further confirmation is necessary, using future CMB maps that may have
  better-controlled systematic errors.
\item For an improved foreground analysis and removal, the foreground maps
  should be pre-decomposed into E and B modes as shown in
  Equation~\ref{equ:basic decomposition}, and the two components should be
  studied separately.
  \label{item:EB-decomposition is necessary}

\end{enumerate}

\Ack{

I sincerely thank Pavel Naselsky, Sebastian von Hausegger and James Creswell
for valuable discussions and suggestions, as well as the anonymous referee for
carefully reading the article and giving very helpful comments. This research
has made use of data/product from the WMAP~\citep{WMAPdata:online} and
Planck~\citep{Planckdata:online} collaborations. Some of the results in this
paper are derived using the HEALPix~\citep{2005ApJ...622..759G} package. This
work was partially funded by the Danish National Research Foundation (DNRF)
through establishment of the Discovery Center and the Villum Fonden through
the Deep Space project. Hao Liu is also supported by the Youth Innovation
Promotion Association, CAS.

}

\appendix

\section{Illustration of the E and B modes in real space}\label{app:EB in real space}

\begin{figure}[!hbtp]
  \centering
  \includegraphics[width=0.48\textwidth]{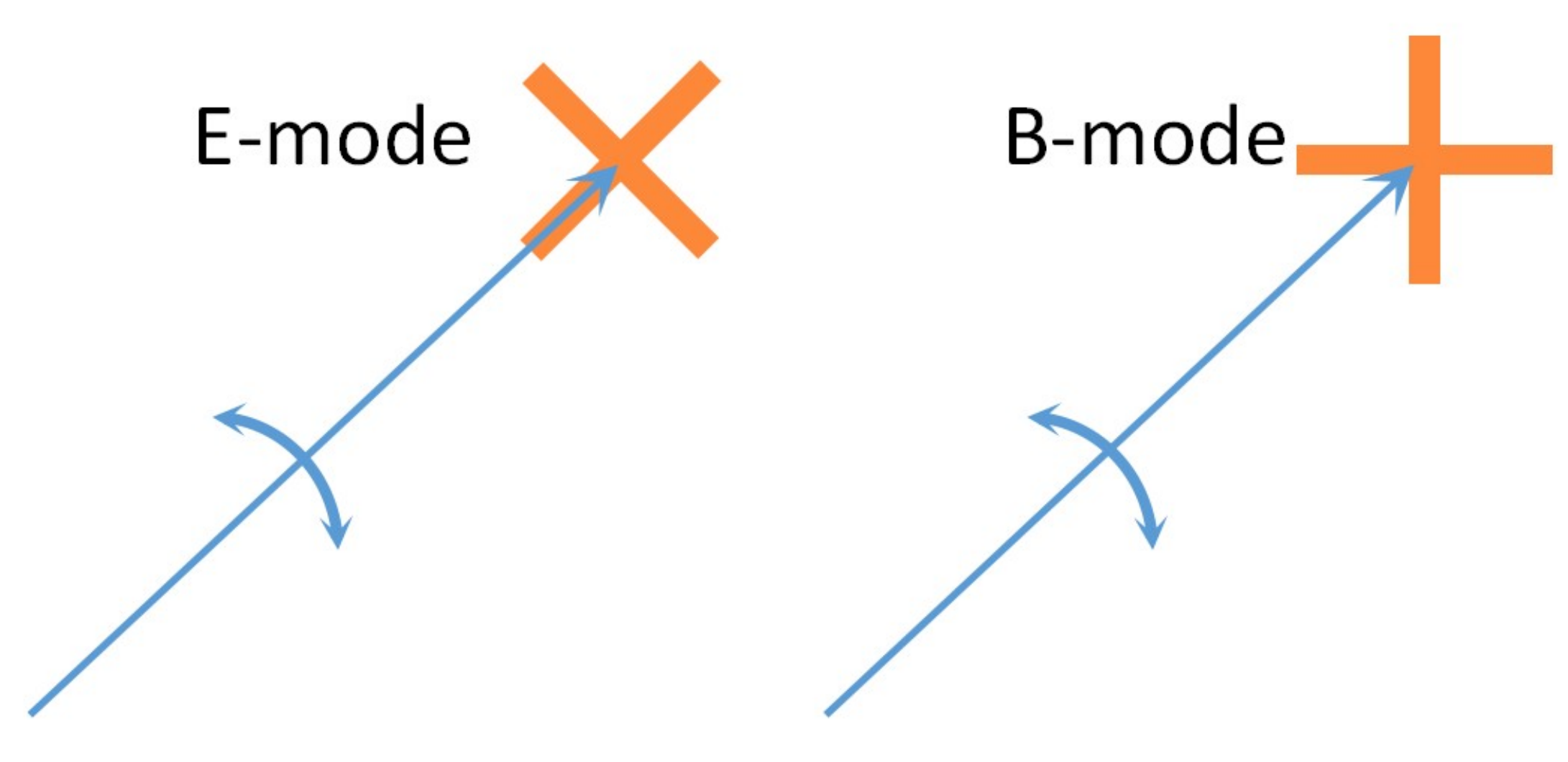}
  \caption{ An illustration of the E and B modes. For each mode, only two
  polarization directions (arms of the cross) are allowed in respect to the
  normal vector (arrow). Only one radial direction is plotted as example, and
  a complete E or B mode is made up of all its rotations. Therefore both E and
  B modes are rotationally invariant by design.}
  \label{fig:EB in real space}
\end{figure}

Analysis of the full sky polarization data requires a prior definition of the
local reference frame for each sky direction, which depends on the choice of
the coordinate system. This is not rotationally invariant and is inconvenient,
so it is usual to decompose the polarized signal into the sum of many circular
structures, with each one being rotationally invariant. For each circular
structure, there are two linearly independent components: E-mode components, in which the polarization direction is either parallel
or
perpendicular to the normal vectors, and
B-mode components, in which the polarization
direction is either $45\degree$ or $135\degree$ from the normal vectors,
as
illustrated in Figure~\ref{fig:EB in real space}.   From the definition, one can
see that the E and B modes are statistically equivalent for a Gaussian random
field, as well as for an ordinary diffusive Galactic foreground. More
information about definitions and practical ways to extract the E and B modes
from CMB maps can be found in~\cite{PhysRevD.55.1830, 1997PhRvD..55.7368K,
1997PhRvL..78.2058K, 0004-637X-503-1-1, 2010A&A...519A.104K,
2016ARA&A..54..227K}

\section{Illustration for the loop magnetic field and projection}\label{app:obs illustration}

Figure~\ref{fig:obs illu} illustrates how the loop magnetic field
affects the polarization observation, in which $\rm{\mathbf{S}}$ is the center
of the supernova remnant, $\rm{\mathbf{P}}$ is one point on the shell,
$\rm{\mathbf{O}}$ is the observer, $\rm{\mathbf{N}}$ is an auxiliary point
above the paper with $\overline{\rm{\mathbf{NP}}}$ being perpendicular to the
paper, and $\overline{\rm{\mathbf{AB}}}$ and $\overline{\rm{\mathbf{XY}}}$ are
auxiliary lines that are perpendicular to the line-of-sight and radial
direction of the supernova, respectively. For point $\rm{\mathbf{P}}$ on the
shell, due to the shell expansion, the magnetic field lines are suppressed and
form along the shell; they should therefore be perpendicular to the normal vector
$\overline{\rm{\mathbf{PS}}}$. This allows two magnetic field components along
$\overline{\rm{\mathbf{PN}}}$ and $\overline{\rm{\mathbf{PY,}}}$ respectively.
The $\overline{\rm{\mathbf{PN}}}$ component is parallel to the tangent
direction of the projected loop on the two-dimensional sky, and is therefore the one discussed in this paper. The $\overline{\rm{\mathbf{PY}}}$ component
can be decomposed into two components along $\overline{\rm{\mathbf{PO}}}$ and
$\overline{\rm{\mathbf{PB,}}}$ respectively.

The background interstellar magnetic field before the SN explosion can have
both regular (smooth) and turbulent components; see for
example~\cite{1996ARA&A..34..155B}. The smooth component is expected to have
only large-scale variation, while the turbulent component is expected to have
random fluctuation at small scales. The threshold for ``large-scale'' and
``small-scale'' is not absolutely defined, but in this work, it is assumed
that the smooth component has negligible variation at the size of an SN
bubble, whereas the turbulent component has random directions in the bubble
volume. In this context, their respective contributions are discussed below.

\subsection{Contribution of the smooth background magnetic field}\label{sub:smooth background mag}

As mentioned above, the smooth component will be suppressed by the supernova,
and the $\overline{\rm{\mathbf{PS}}}$ component will be erased. The
$\overline{\rm{\mathbf{PO}}}$ component is therefore aligned with the line-of-sight
(LOS), which can not generate any visible polarization (but is related to the
Faraday rotation). The contribution of the $\overline{\rm{\mathbf{PB}}}$ component is
projected from the $\overline{\rm{\mathbf{PY}}}$ component by a factor of
$\sin(\theta)$, which is already suppressed. Furthermore, since the LOS will
cross a shell twice, at $\rm{\mathbf{P}}$ and $\rm{\mathbf{P}}'$, the
$\overline{\rm{\mathbf{PB}}}$ components at these two points will cancel each
other out, which means a further cancellation. Therefore, after projection and
LOS integration, the major component of the magnetic field that is effective
for polarization is the $\overline{\rm{\mathbf{PN}}}$ component. However, the
$\overline{\rm{\mathbf{PB}}}$ component after suppression and cancellation can
be small but non-zero, which might be a source of the B-mode emission from
loops.
\begin{figure}[!htb]
  \centering
  \includegraphics[width=0.48\textwidth]{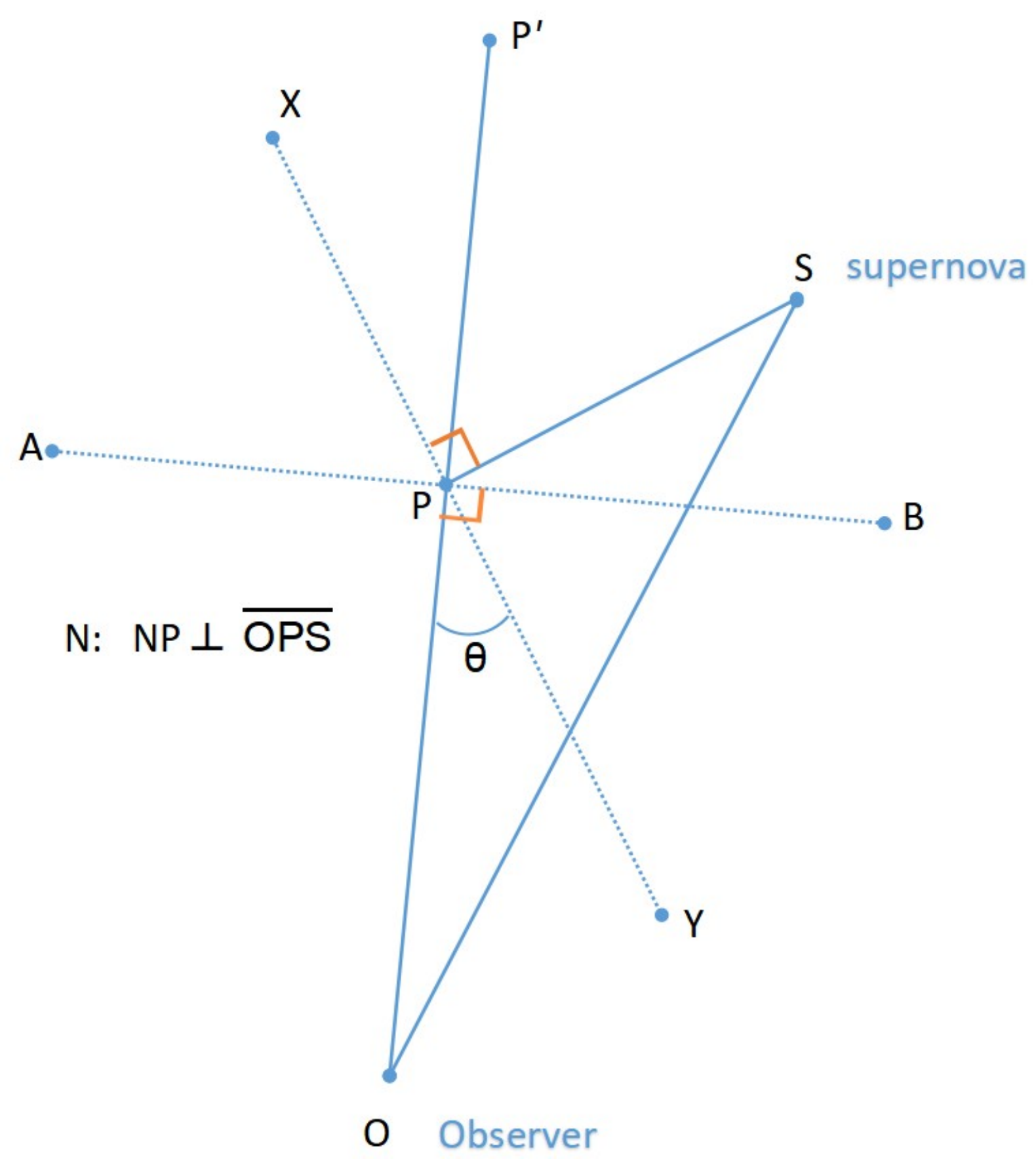}
  \caption{ Illustration on how the polarized loop signal is projected.
  $\rm{\mathbf{S}}$ is the supernova, $\rm{\mathbf{O}}$ is the observer,
  $\rm{\mathbf{P}}$ is one point on the shell, $\overline{\rm{\mathbf{AB}}}$
  is perpendicular to $\overline{\rm{\mathbf{OP}}}$, and
  $\overline{\rm{\mathbf{XY}}}$ is perpendicular to
  $\overline{\rm{\mathbf{SP}}}$. $\rm{\mathbf{N}}$ lies above the paper and
  $\overline{\rm{\mathbf{NP}}}$ is perpendicular to the
  $\overline{\rm{\mathbf{OPS}}}$ plane.}
  \label{fig:obs illu}
\end{figure}

A complete calculation of the LOS integration in case of smooth background
magnetic field is very difficult. Therefore, a much-simplified toy
model is provided for example, with the following assumptions/simplifications:
\begin{enumerate}
\item The observer stays at $[x,y,z]=[0,0,0]$, the center of SN is
        $[0,0,1]$, and the coordinate of $\rm{\mathbf{P}}$ is
        $[R\sin(\theta)\cos(\phi), R\sin(\theta)\sin(\phi), R\cos(\theta)]$.
        \label{item:coord}
\item The SN will only erase the $\overline{\rm{\mathbf{PS}}}$ component.
\item The background magnetic field (before SN explosion) is $\rm{\mathbf{B}}$
  = $[\sin(\theta_1)\cos(\phi_1)$, $\sin(\theta_1)\sin(\phi_1)$,
  $\cos(\theta_1)]$, whose direction is uniformly distributed on the sphere.
  However, for one realization, $\rm{\mathbf{B}}$ is a constant vector.
\end{enumerate}

Therefore, the effective magnetic field along $\overline{\rm{\mathbf{PN}}}$ and
$\overline{\rm{\mathbf{PB}}}$ at point $\rm{\mathbf{P}}$ are:
\begin{eqnarray}\label{equ:B comp smooth}
B_{PN} &=& \sin(\theta_1)\sin(\delta) \\ \nonumber
B_{PB} &=& \frac{R-\cos(\theta)}{1+R^2-2R\cos(\theta)}[R\cos(\theta_1)\sin(\theta)+ \\ \nonumber
& & (1-R\cos(\theta))\cos(\delta)\sin(\theta_1)],
\end{eqnarray}
where $\delta = \phi_1-\phi$.

The polarization angle is bound with the magnetic field, and so are the Stokes
parameters. Therefore, Equation~\ref{equ:B comp smooth} gives the following
Stokes parameters:
\begin{eqnarray}
Q(R,\theta,\theta_1,\delta) &=& \cos[2\arctan(B_{PN},B_{PB})] \\ \nonumber 
U(R,\theta,\theta_1,\delta) &=& \sin[2\arctan(B_{PN},B_{PB})].
\end{eqnarray}
Here, according to the coordinate system shown in Figure~\ref{fig:obs illu},
the Q Stokes parameter corresponds to the E-mode, and the U Stokes parameter
corresponds to the B-mode. 

Apparently, if $B_{PN}$ or $B_{PB}$ is zero, then
$U(R,\theta,\theta_1,\delta)\equiv 0$, and the signal is pure E-mode. Since
$B_{PN}$ is a constant in the LOS integration, this provides a good example:
if the smooth background magnetic field is along $\overline{\rm{\mathbf{OS}}}$
(so $\theta_1=0$), then the final polarization signal is always pure E-mode.

For an LOS integration, the amplitude of the $(Q,U)$ Stokes parameters is
assumed to be scaled by
\begin{eqnarray}
f(R,\theta,\gamma) &=& \frac{1}{(1+R^2-2R\cos(\theta))^\gamma R^2},
\end{eqnarray}
where $1+R^2-2R\cos(\theta)$ is the length of $\overline{\rm{\mathbf{PS}}}$,
and $\gamma\ge0$ is a free parameter that describes the decay of the SN signal
according to the shell radius (this is a simplified assumption). Therefore,
the integrated stoke parameters along the LOS are:
\begin{eqnarray}\label{equ:qu_LOS_int}
Q_{s}(\theta,\theta_1,\delta)&=&\int_0^{r}Q(R,\theta,\theta_1,\delta)f(R,\theta,\gamma) dR \\ \nonumber
U_{s}(\theta,\theta_1,\delta)&=&\int_0^{r}U(R,\theta,\theta_1,\delta)f(R,\theta,\gamma) dR \\ \nonumber
\rho(\theta,\theta_1,\delta)&=& \frac{Q_{s}(\theta,\theta_1,\delta)}{U_{s}(\theta,\theta_1,\delta)}.
\end{eqnarray}

With Equation~\ref{equ:qu_LOS_int} and various values of $\theta$, $\theta_1$,
$\delta$, one can calculate the ratio $\rho(\theta,\theta_1,\delta)$ between
the two components, which is exactly the EB-ratio at the given sky direction
after LOS integration. With various parameters this ratio can either be
greater or smaller than 1. For statistical purposes, uniformly distributed sky
directions are assumed for the LOS, the initial magnetic field runs
$10^4$ simulations, and the median EB-ratio is roughly 1.7, which moderately
prefers the E-mode.

\begin{figure*}[!htb]
  \centering
  \includegraphics[width=0.32\textwidth]{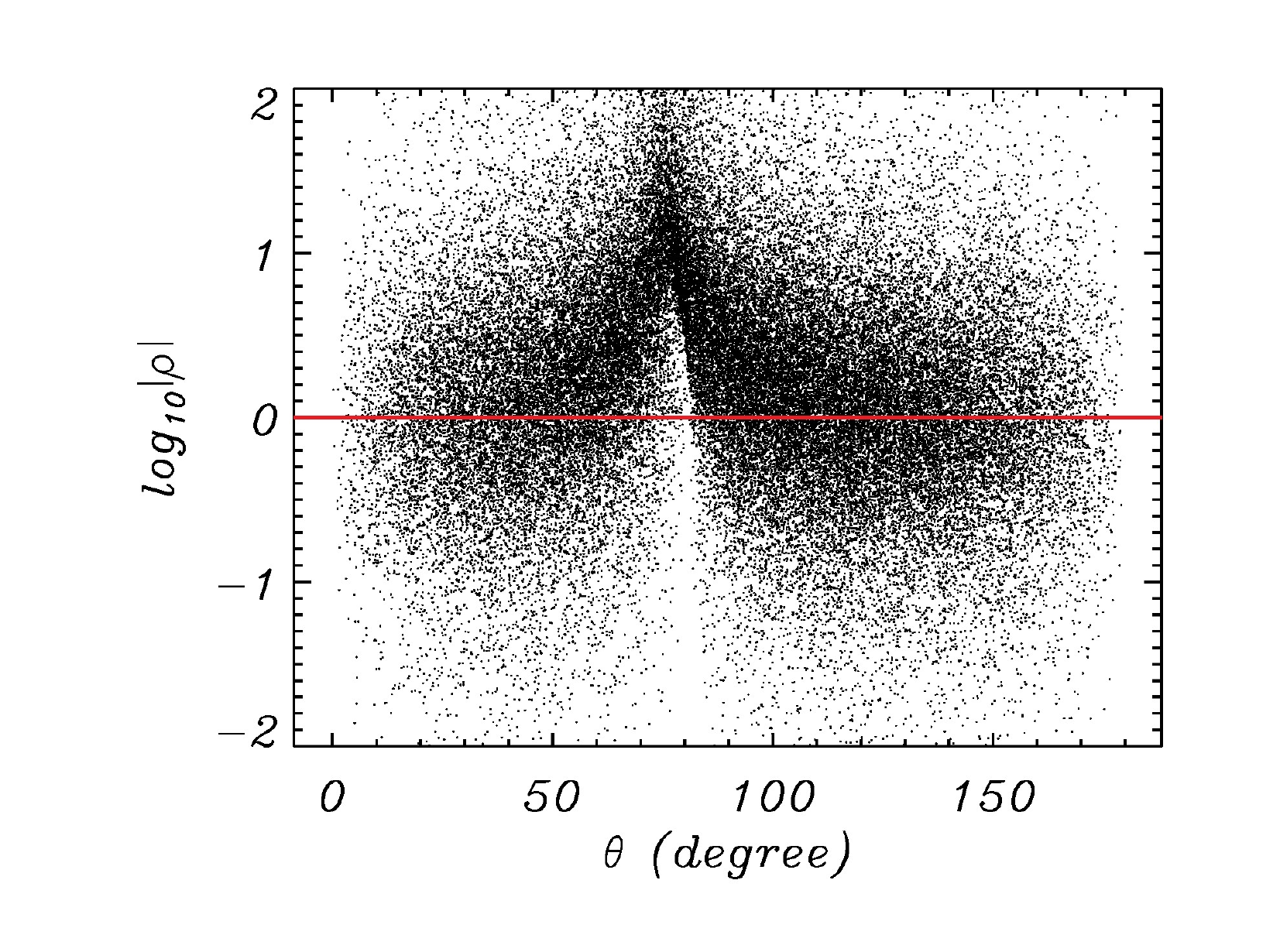}
  \includegraphics[width=0.32\textwidth]{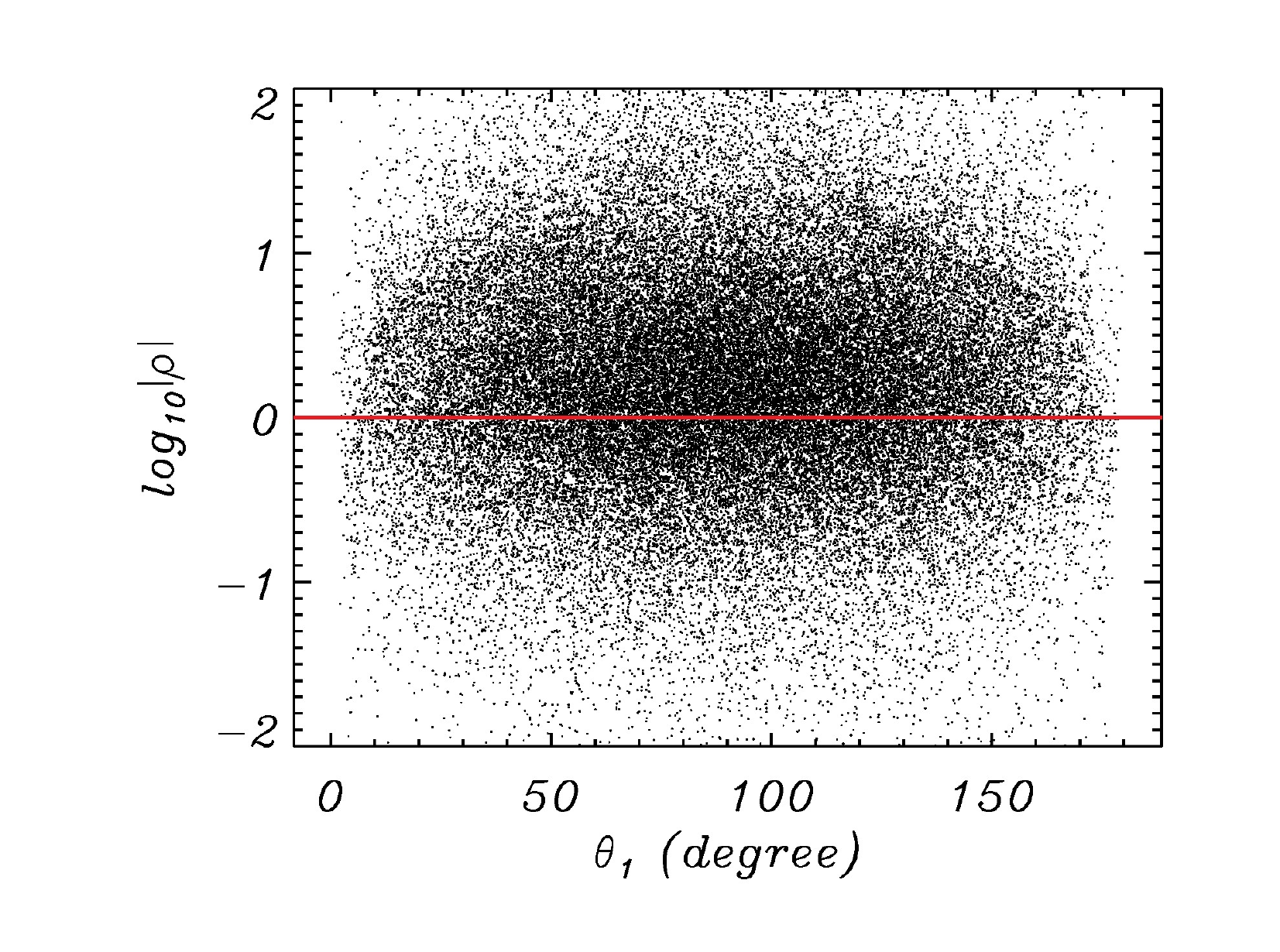}
  \includegraphics[width=0.32\textwidth]{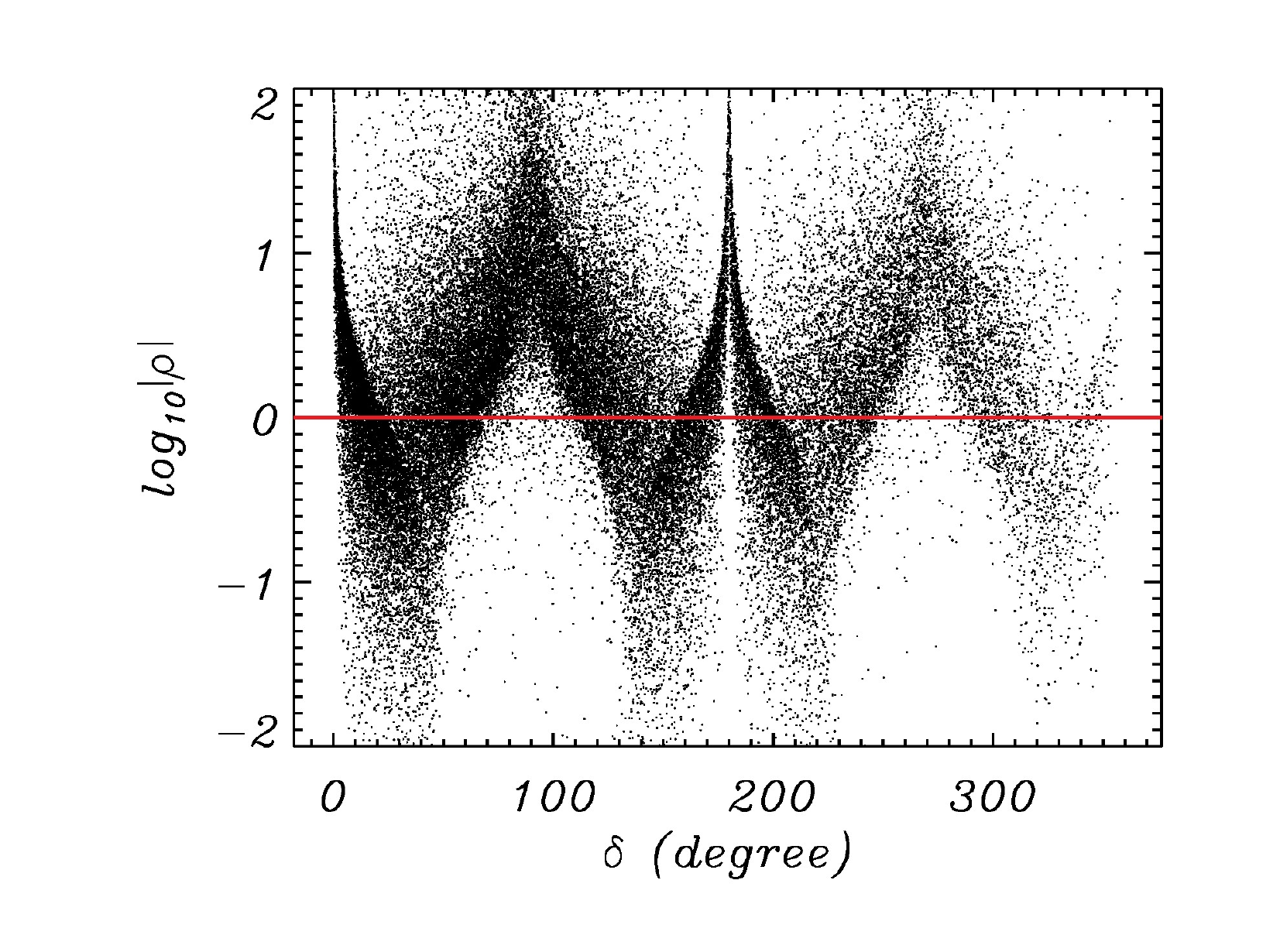}
  \caption{ Logarithm EB-ratio $\log_{10}(|\rho|)$ in case of smooth
        background magnetic field, and as function of $\theta$ (left), $\theta_1$
        (middle) and $\delta$ (right). The unplotted parameters are random for
        each panel. The red lines are for $\rho=1$, so the dots above the red line
        have higher E-mode, and the dots below the red line have higher B-mode.
        I note that the results are not sensitive to $\gamma$.}
  \label{fig:loop LOS model}
\end{figure*}

To roughly illustrate the expected EB-ratio $\rho(\theta,\theta_1,\delta)$ for
various choices of the parameters, $\log_{10}(|\rho|)$ is plotted as function
of $\theta$, $\theta_1$ and $\delta$ in Figure~\ref{fig:loop LOS model}, and
in each panel the unplotted parameters are simply random. I note that the
results are not sensitive to $\gamma$. One can see that the E-mode is
apparently preferred for $\theta$ around $75\degree$, and is moderately
preferred for all $\theta_1$. There is also significant sinusoidal modulation
associated with the $\delta$ parameter with peaks around
$\delta=0\degree,\,90\degree,\,180\degree\dots$ However, it must be emphasized
that Figure~\ref{fig:loop LOS model} is only statistical, which cannot cast
effective constraints on a specific realization.

\subsection{Contribution of the turbulent background magnetic field}\label{sub:turbulence background mag}

The turbulent magnetic field is expected to have random directions for
different parts of the sky; however, due to the supernova explosion, the
magnetic field components along the $\overline{\rm{\mathbf{PS}}}$ direction
are erased, and the magnetic field is constrained within the
$\overline{\rm{\mathbf{NPY}}}$ plane with uniformly distributed directions,
which can be represented in a circular form:
\begin{eqnarray}\label{equ:turbulence B circular}
u_0 &=&|B|\sin(\phi) \\ \nonumber v_0 &=&|B|\cos(\phi),
\end{eqnarray}
where $u_0$ is the magnetic field component along the
$\overline{\rm{\mathbf{PN}}}$ direction, $v_0$ is the component along the
$\overline{\rm{\mathbf{PY}}}$ direction, $\phi$ is the random phase angle in
the $\overline{\rm{\mathbf{NPY}}}$ plane, and $|B|$ is the amplitude of the
magnetic field.

For the region around $\rm{\mathbf{P}}$, the contribution of the turbulent
magnetic field can be regarded as the integration of
Equation~\ref{equ:turbulence B circular} with uniformly distributed $\phi$.
Considering the projection from the $\overline{\rm{\mathbf{NPY}}}$ plane to
the $\overline{\rm{\mathbf{NPB}}}$ plane that is perpendicular to the LOS,
Equation~\ref{equ:turbulence B circular} changes from circular to elliptical:
\begin{eqnarray}
u&=&|B|\sin(\psi) \\ \nonumber
v&=&|B|\cos(\psi)\sin(\theta) \\ \nonumber
|B'|&=&|B|\sqrt{\sin^2(\psi)+\cos^2(\psi)\sin^2(\theta),}
\end{eqnarray}
where $u$ is the magnetic field component along the
$\overline{\rm{\mathbf{PN}}}$ direction, $v$ is the component along the
$\overline{\rm{\mathbf{PB}}}$ direction, $\psi$ is the phase angle of the
polar system in the $\overline{\rm{\mathbf{NPB}}}$ plane, and $|B'|$ is the
amplitude of the effective magnetic field after projection. The polarization
angle is simply $\psi'=\psi+90\degree$; therefore, for the synchrotron
emission, the integrations of the Q, U Stokes parameters for the region around
$\rm{\mathbf{P}}$ are:
\begin{eqnarray}\label{equ:QU int ellipse}
Q&\propto&\int_0^{2\pi}|B'|^2\cos(2\psi')d\psi' \, \propto \, \cos^2(\theta)\\ \nonumber
U&\propto&\int_0^{2\pi}|B'|^2\sin(2\psi')d\psi'=0.
\end{eqnarray}
As mentioned already in Section~\ref{sub:smooth background mag}, here the Q
Stokes parameter represents the E-mode, and the U-Stokes parameter represents
the B-mode. Therefore, one can see that the turbulent magnetic field always
gives pure E-mode.

The physical meaning of Equation~\ref{equ:QU int ellipse} is: due to the
``$2\psi$'' rule of the polarization angles, the symmetry of the integral
along the ellipse is broken, so the two directions separated by $\pi$ no
longer cancel each other, which is the source of non-zero E-mode. If $2\psi'$
is replaced by $\psi'$ in Equation~\ref{equ:QU int ellipse}, then both
integrals are zero.

With Equation~\ref{equ:QU int ellipse}, the conclusion is apparent: supernova
explosion plus a turbulent background magnetic field will produce pure E-mode
in the polarized signal. However, it is always possible that, due to asymmetry
of the supernova explosion, the B-mode is small but non-zero.

\section{Note added to proofs -- updated table with the Planck final data release}
During the production stage of this paper, Planck released its final results
(\url{https://www.cosmos.esa.int/web/planck/publications}), which is used to
update Table~\ref{tab:mad}. The new results are listed in
Table~\ref{tab:mad1}. The Planck final data release is expected to have lower
systematics, and one can indeed see that the blue regions in
Table~\ref{tab:mad}, which were marked as suspicious due to systematics,
become apparently lower in Table~\ref{tab:mad1}, which makes the conclusions
in this work more robust.
\begin{table*}[!htb]
 \caption{Same to Table~\ref{tab:mad}, but the Planck results are updated with
 the Planck final data release. The blue regions in Table~\ref{tab:mad} that
 were marked as suspicious due to systematic become apparently lower in this
 table, indicating a better support to the conclusions in this work.}
 \centering
 \begin{tabular}{|c|c|c|c|c|c|c|c|c|c|c|c|c|} \hline
  \rowcolor{gray!40!}
  Band      &      K &     30                 &     Ka &      Q &     44                    &      V &   70   &      W &    100 &    143 &    217 &    353 \\ \hline
  \rowcolor{gray!40!}
  $\nu$ (GHz) &        
\emph{  22.8} & \emph{  28.4} & \emph{  33.0} & \emph{  40.7} & \emph{  44.1} & \emph{  60.8} &
\emph{  70.4} & \emph{  93.5} & \emph{ 100.0} & \emph{ 143.0} & \emph{ 217.0} & \emph{ 353.0} \\ \hline
 In \& Out &  15.6 &   \cellcolor{blue!40}18.9 &   16.5 &   16.7 &   \cellcolor{blue!40}23.0 &   29.6 &   29.9 &   22.0 &   22.4 &   19.0 &   17.3 &   17.1 \\ \hline
 Inside    &  16.2 &   \cellcolor{blue!40}18.8 &   16.6 &   16.4 &   \cellcolor{blue!40}20.5 &   27.4 &   27.4 &   22.2 &   19.1 &   19.3 &   18.0 &   17.9 \\ \hline 
 Outside   &  14.3 &   \cellcolor{blue!40}19.2 &   16.2 &   17.3 &   \cellcolor{blue!40}28.3 &   34.5 &   34.4 &   21.7 &   30.3 &   18.3 &   15.6 &   15.1 \\ \hline
 Bicep2    &   9.7 &   \cellcolor{blue!40}13.9 &    9.0 &    6.8 &   \cellcolor{blue!40}17.0 &   21.8 &   38.1 &   19.5 &    \cellcolor{blue!40}8.7 &    7.8 &    5.6 &    5.2 \\ \hline
 \end{tabular}
 \label{tab:mad1}
\end{table*}

\end{document}